\documentclass[%
 reprint,
 showpacs,
 showkeys,
 preprintnumbers,
 amsmath,amssymb,
 aps,
 prl,
  longbibliography,
 ]{revtex4-1}

\usepackage[breaklinks=true,colorlinks=true,anchorcolor=blue,citecolor=blue,filecolor=blue,menucolor=blue,pagecolor=blue,urlcolor=blue,linkcolor=blue]{hyperref}
\usepackage{graphicx}% Include figure files
\usepackage{xcolor}

\begin{document}

\title{How much contextuality?}

\author{Karl Svozil}
\affiliation{Institute of Theoretical Physics, Vienna
    University of Technology, Wiedner Hauptstra\ss e 8-10/136, A-1040
    Vienna, Austria}
\email{svozil@tuwien.ac.at} \homepage[]{http://tph.tuwien.ac.at/~svozil}

\date{\today}

\begin{abstract}
The amount of contextuality is quantified in terms of the probability of the necessary violations of noncontextual assignments to counterfactual elements of physical reality.
\end{abstract}

%\preprint{CDMTCS preprint nr. 372/2009}
\maketitle

Some of the mind boggling features attributed to quantized systems are their alleged ability
to counterfactually~\citep{svozil-2006-omni,vaidman:2009} respond to complementary queries~\cite{epr,clauser},
as well as their capacity to experimentally render outcomes
which have not been encoded into them prior to measurement~\cite{zeil-99}.
Moreover, under certain ``reasonable'' assumptions, and by excluding various exotic quasi-classical possibilities~\cite{pitowsky-82,meyer:99},
quantum mechanics appears to ``outperform'' classical correlations by allowing
higher-that-classical coincidences of certain events,
reflected by violations of Boole-Bell type constraints on classical probabilities~\cite{Boole-62,froissart-81,pitowsky}.
One of the unresolved issues is the reason (beyond geometric and formal arguments)
for the quantitative form of these violations~\cite{cirelson,filipp-svo-04-qpoly-prl};
in particular, why Nature should not allow higher-than-quantum or maximal violations~\cite{pop-rohr,svozil-krenn} of
Boole's conditions of possible experience~\cite[p.~229]{Boole-62}.

The Kochen-Specker theorem~\cite{kochen1}, stating the impossibility of a consistent truth assignment to the potential outcomes of
(even a finite number) of interlinked complementary observables,
gave further indication for the absence of classical omniscience in the quantum domain.
One possibility to interpret these findings, and the prevalent one among physicists,
is in terms of contextuality:
it is thereby implicitly assumed that all potentially observable elements of physical reality~\cite{epr} exist prior to any measurement; albeit
any such (potential) measurement outcome (the entirety of which could thus consistently pre-exist before the actual measurement)
depends on whatever other observables (the context) are co-measured alongside~\cite{bohr-1949,bell-66}.
As, contrary to a very general interpretation of that assumption, the quantum mechanical observables are represented context independently,
any such contextual behavior should be restricted to {\em single} quanta and outcomes within the quantum statistical bounds.

Einstein-Podolsky-Rosen type experiments~\cite{epr}
for entangled higher than two-dimensional quantized systems seem to indicate that
contextuality, if viable, will remain hidden to any direct physical operationalization
(and thus might be criticized to be metaphysical)
even if counterfactual measurements are allowed~\cite{svozil:040102}.
Because ``the
immense majority of the experimental violations of Bell
inequalities does not prove quantum nonlocality, but just
quantum contextuality''~\cite{cabello:210401},
current claims of proofs of noncontextuality are solely based on violations of classical constraints in Boole-Bell-type,
Kochen-Specker-type, or Greenbergerger-Horne-Zeilinger-type  configurations.

Nevertheless, insistence on the simultaneous physical contextual coexistence
of certain finite sets of counterfactual observables results in truth assignments which could be explicitly illustrated by
a {\em forced} tabulation \cite{peres222,svozil_2010-pc09}
of contextual truth values for Boole-Bell-type or Kochen-Specker-type configurations.
Here contextual means that the truth value of a particular quantum observable depends
on whatever other observables are measured alongside this particular observable.
Any forced tabulation of truth values would render occurrences of mutually contradicting
outcomes of truth or falsity of one and the same observable, depending on the measurement context~\cite{svozil-2008-ql}.
The amount of this violation of noncontextuality can be quantified by the frequency of occurrence of contextuality.
In what follows these frequencies will be calculated for a number of experimental configurations suggested in the literature.

First, consider the generalized Clauser-Horne-Shimony-Holt (CHSH) inequality
\begin{equation}
-\lambda \le E(a,b)+E(a,b')+E(a',b)-E(a',b') \le \lambda
\label{2011-enough-e1}
\end{equation}
which, for $\lambda =2$ and $\lambda =2\sqrt{2}$,
represents bounds for classical~\cite{chsh,clauser} and quantum~\cite{cirelson:80}
expectations of dichotomic observables with outcomes ``$-1$'' and ``$+1$,'' respectively.
The algebraically maximal violation associated with $\lambda =4$
is attainable only for hypothetical ``nonlocal boxes''~\cite{pop-rohr,svozil-krenn,popescu-97,PhysRevA.71.022101}
or by bit exchange~\cite{svozil-2004-brainteaser}.

Eq.~(\ref{2011-enough-e1}) can be rewritten in an explicitly contextual form by the substitution
\begin{equation}
E(x,y)
\mapsto
E(x_y,y_x)
,
\label{2011-enough-e2}
\end{equation}
where $x_y$ stands for ``observable $x$ measured alongside observable $y$''
\cite{svozil_2010-pc09}.
Contextuality manifests itself through $x_y\neq x_{y'}$.
Because in the particular CHSH configuration there are no other observables measured alongside the ones
that appear already in Eq.~(\ref{2011-enough-e1}), this form is without ambiguity.

For the sake of simplicity, suppose one would like to force the algebraic maximum of $\lambda = 4$
upon  Eq.~(\ref{2011-enough-e1}), and suppose that only one observable, say $b'$,
is contextual (a highly counterintuitive assumption).
Then one obtains
\begin{equation}
(\pm 1) (\pm 1)
+
(\pm 1) x
+
(\pm 1) (\pm 1)
-
(\pm 1) (-x)
=
4,
\label{2011-enough-e3}
\end{equation}
and thus $x=\pm 1$.
Thus, in order to reach the algebraic maximum, contextuality has to be maximal,
that is $b'_a = -b'_{a'}$ for any quantum.
Table~\ref{2011-enough-t1} enumerates the two possible truth value assignments
associated with this configuration.
\begin{table}
\begin{center}
\begin{tabular}{ccccccccccccccccccccccccccccccccc}
\hline\hline
$a_b$&$a_{b'}$&$a'_b$&$a'_{b'}$&$b_a$&$b_{a'}$&$b'_a$&$b'_{a'}$\\
\hline
$+1$&$+1$&$+1$&$+1$&$+1$&$+1$&$+1$&$-1 $\\
$-1$&$-1$&$-1$&$-1$&$-1$&$-1$&$-1$&$+1 $\\
\hline
$+1$&$+1$&$+1$&$+1$&$+1$&$+1$&$+1$&$+1 $\\
$-1$&$-1$&$-1$&$-1$&$-1$&$-1$&$-1$&$-1 $\\
\hline\hline
\end{tabular}
\end{center}
\caption{The first two rows represent contextual assignments associated with an algebraic maximal rendition ($\lambda =4$)
of the CHSH inequality.
The third and the fourth assignments are noncontextual.}
\label{2011-enough-t1}
\end{table}

That contextuality could accommodate any bound $0< \lambda <4$
can be demonstrated by  interpreting
all possible noncontextual and contextual assignments, as well as the resulting
corresponding joint expectations enumerated in Table~\ref{2011-enough-t2}
as vertices of a convex correlation polytope.
\begin{table*}
\begin{center}
\begin{tabular}{cccccccc|ccccccccccccccccccccccccc}
\hline\hline
$a_b$&$a_{b'}$&$a'_b$&$a'_{b'}$&$b_a$&$b_{a'}$&$b'_a$&$b'_{a'}$&$a_b b_a$&$a_{b'} b'_a$&$a'_b b_{a'}$&$a'_{b'} b'_{a'}$     \\
\hline
$  -1$&$   -1$&$   -1$&$   -1$&$   -1$&$   -1$&$   -1$&$   -1$&$   +1$&$   +1$&$   +1$&$   +1$\\
$  -1$&$   -1$&$   -1$&$   -1$&$   -1$&$   -1$&{\color{green}$\bf{}-1$}&{\color{red}$\bf{}+1$}&$   +1$&$   +1$&$   +1$&$   -1$\\
$  -1$&$   -1$&$   -1$&$   -1$&$   -1$&$   -1$&{\color{red}$\bf{}+1$}&{\color{green}$\bf{}-1$}&$   +1$&$   -1$&$   +1$&$   +1$\\
$  -1$&$   -1$&$   -1$&$   -1$&$   -1$&$   -1$&$   +1$&$   +1$&$   +1$&$   -1$&$   +1$&$   -1$\\
$  -1$&$   -1$&$   -1$&$   -1$&{\color{green}$\bf{}-1$}&{\color{red}$\bf{}+1$}&$   -1$&$   -1$&$   +1$&$   +1$&$   -1$&$   +1$\\
$  -1$&$   -1$&$   -1$&$   -1$&{\color{green}$\bf{}-1$}&{\color{red}$\bf{}+1$}&{\color{green}$\bf{}-1$}&{\color{red}$\bf{}+1$}&$   +1$&$   +1$&$   -1$&$   -1$\\
$  -1$&$   -1$&$   -1$&$   -1$&{\color{green}$\bf{}-1$}&{\color{red}$\bf{}+1$}&{\color{red}$\bf{}+1$}&{\color{green}$\bf{}-1$}&$   +1$&$   -1$&$   -1$&$   +1$\\
$  -1$&$   -1$&$   -1$&$   -1$&{\color{green}$\bf{}-1$}&{\color{red}$\bf{}+1$}&$   +1$&$   +1$&$   +1$&$   -1$&$   -1$&$   -1$\\
 $\vdots$&$\vdots$&$\vdots$&$\vdots$&$\vdots$&$\vdots$&$\vdots$&$\vdots$&$\vdots$&$\vdots$&$\vdots$&$\vdots$\\
%$  -1$&$   +1$&$   -1$&$   +1$&$   -1$&$   +1$&$   -1$&$   -1$&$   +1$&$   -1$&$   -1$&$   -1$\\
 {\color{green}$\bf{}-1$}&{\color{red}$\bf{}+1$}&{\color{green}$\bf{}-1$}&{\color{red}$\bf{}+1$}&{\color{green}$\bf{}-1$}&{\color{red}$\bf{}+1$}&{\color{green}$\bf{}-1$}&{\color{red}$\bf{}+1$}&$   +1$&$   -1$&$   -1$&$   +1$\\
 $\vdots$&$\vdots$&$\vdots$&$\vdots$&$\vdots$&$\vdots$&$\vdots$&$\vdots$&$\vdots$&$\vdots$&$\vdots$&$\vdots$\\
 {\color{red}$\bf{}+1$}&{\color{green}$\bf{}-1$}&{\color{red}$\bf{}+1$}&{\color{green}$\bf{}-1$}&{\color{red}$\bf{}+1$}&{\color{green}$\bf{}-1$}&{\color{red}$\bf{}+1$}&{\color{green}$\bf{}-1$}&$   +1$&$   -1$&$   -1$&$   +1$\\
 $\vdots$&$\vdots$&$\vdots$&$\vdots$&$\vdots$&$\vdots$&$\vdots$&$\vdots$&$\vdots$&$\vdots$&$\vdots$&$\vdots$\\
 $  +1$&$   +1$&$   +1$&$   +1$&$   +1$&$   +1$&{\color{green}$\bf{}-1$}&{\color{red}$\bf{}+1$}&$   +1$&$   -1$&$   +1$&$   +1$\\
 $  +1$&$   +1$&$   +1$&$   +1$&$   +1$&$   +1$&{\color{red}$\bf{}+1$}&{\color{green}$\bf{}-1$}&$  +1$&$   +1$&$   +1$&$   -1$\\
 $  +1$&$   +1$&$   +1$&$   +1$&$   +1$&$   +1$&$   +1$&$   +1$&$   +1$&$   +1$&$   +1$&$   +1$\\
\hline\hline
\end{tabular}
\end{center}
\caption{(Color online) Contextual (bold) and noncontextual value assignments, and the associated joint values.}
\label{2011-enough-t2}
\end{table*}
According to the  Minkoswki-Weyl representation theorem~\cite[p~29]{ziegler}, an equivalent representation of the
associated convex polyhedron is in terms of the halfspaces defined by Boole-Bell type inequalities of the form
\begin{equation}
\begin{array}{l}
-1\le E(a_b)+E(b_a)+E(a_b b_a),  \\
-1\le  E(a_b)-E(b_a)-E(a_b b_a),  \\
-1\le  -E(a_b)+E(b_a)-E(a_b b_a),  \\
-1\le  -E(a_b)-E(b_a)+E(a_b b_a),
\end{array}
\label{2011-enough-e4}
\end{equation}
(and the inequalities resulting from permuting
$a \leftrightarrow a'$,
$b \leftrightarrow b'$)
which, for $E(a_b)=E(b_a)=0$, reduce to $-1\le E(a_b b_a)\le 1$.
Note that, by taking only the 16 context-independent ($x_y=x_{y'}$) from all the 256 assignments,
the CHSH inequality~(\ref{2011-enough-e1}) with $\lambda =2$ is recovered.

Next, for the sake of demonstration, an example configuration will be given that conforms to
Tsirel'son's maximal quantum bound of $\lambda = 2\sqrt{2}$~\cite{cirelson}.
Substituting this for $2\sqrt{2}$ in Eq.~(\ref{2011-enough-e3})
yields $x=\pm (\sqrt{2}-1)$; that is,
the (limit) frequency for the occurrence of contextual assignments
$
b'_{a} = - b'_{a'}
$
as enumerated in
Table~\ref{2011-enough-t1} with respect to the associated noncontextual assignments
$
b'_{a} = b'_{a'}
$
(rendering 2 to the sum of terms in the CHSH expression) should be
$(\sqrt{2}-1) : (2-\sqrt{2})$.
More explicitly,
if there are four different assignments,
enumerated in Table~\ref{2011-enough-t1},
which may contribute quantum mechanically by the correct (limiting) frequency,
then Table~\ref{2011-enough-t3} is a simulation of 20 assignments rendering the maximal quantum bound
for the CHSH inequalities.
\begin{table}
\begin{center}
\begin{tabular}{cccccccc} % |ccccccccccccccccccccccccc}
\hline\hline
$a_b$&$a_{b'}$&$a'_b$&$a'_{b'}$&$b_a$&$b_{a'}$&$b'_a$&$b'_{a'}$\\% &$a_b b_a$&$a_{b'} b'_a$&$a'_b b_{a'}$&$a'_{b'} b'_{a'}$     \\
\hline
$+1$&$+1$&$+1$&$+1$&$+1$&$+1$&$+1$&$+1$\\ % &$+1$&$+1$&$+1$&$+1$\\
$-1$&$-1$&$-1$&$-1$&$-1$&$-1$&$-1$&$-1$\\ % &$+1$&$+1$&$+1$&$+1$\\
$+1$&$+1$&$+1$&$+1$&$+1$&$+1$&{\color{green}$\bf{}+1$}&{\color{red}$\bf{}-1$}\\ % &$+1$&$+1$&$+1$&$-1$\\
$-1$&$-1$&$-1$&$-1$&$-1$&$-1$&{\color{green}$\bf{}-1$}&{\color{red}$\bf{}+1$}\\ % &$+1$&$+1$&$+1$&$-1$\\
$-1$&$-1$&$-1$&$-1$&$-1$&$-1$&$-1$&$-1$\\ % &$+1$&$+1$&$+1$&$+1$\\
$+1$&$+1$&$+1$&$+1$&$+1$&$+1$&{\color{green}$\bf{}+1$}&{\color{red}$\bf{}-1$}\\ % &$+1$&$+1$&$+1$&$-1$\\
$+1$&$+1$&$+1$&$+1$&$+1$&$+1$&{\color{green}$\bf{}+1$}&{\color{red}$\bf{}-1$}\\ % &$+1$&$+1$&$+1$&$-1$\\
$+1$&$+1$&$+1$&$+1$&$+1$&$+1$&$+1$&$+1$\\ % &$+1$&$+1$&$+1$&$+1$\\
$-1$&$-1$&$-1$&$-1$&$-1$&$-1$&{\color{green}$\bf{}-1$}&{\color{red}$\bf{}+1$}\\ % &$+1$&$+1$&$+1$&$-1$\\
$+1$&$+1$&$+1$&$+1$&$+1$&$+1$&$+1$&$+1$\\ % &$+1$&$+1$&$+1$&$+1$\\
$-1$&$-1$&$-1$&$-1$&$-1$&$-1$&$-1$&$-1$\\ % &$+1$&$+1$&$+1$&$+1$\\
$+1$&$+1$&$+1$&$+1$&$+1$&$+1$&$+1$&$+1$\\ % &$+1$&$+1$&$+1$&$+1$\\
$+1$&$+1$&$+1$&$+1$&$+1$&$+1$&$+1$&$+1$\\ % &$+1$&$+1$&$+1$&$+1$\\
$+1$&$+1$&$+1$&$+1$&$+1$&$+1$&{\color{green}$\bf{}+1$}&{\color{red}$\bf{}-1$}\\ % &$+1$&$+1$&$+1$&$-1$\\
$+1$&$+1$&$+1$&$+1$&$+1$&$+1$&{\color{green}$\bf{}+1$}&{\color{red}$\bf{}-1$}\\ % &$+1$&$+1$&$+1$&$-1$\\
$+1$&$+1$&$+1$&$+1$&$+1$&$+1$&$+1$&$+1$\\ % &$+1$&$+1$&$+1$&$+1$\\
$-1$&$-1$&$-1$&$-1$&$-1$&$-1$&{\color{green}$\bf{}-1$}&{\color{red}$\bf{}+1$}\\ % &$+1$&$+1$&$+1$&$-1$\\
$-1$&$-1$&$-1$&$-1$&$-1$&$-1$&$-1$&$-1$\\ % &$+1$&$+1$&$+1$&$+1$\\
$+1$&$+1$&$+1$&$+1$&$+1$&$+1$&{\color{green}$\bf{}+1$}&{\color{red}$\bf{}-1$}\\ % &$+1$&$+1$&$+1$&$-1$\\
$+1$&$+1$&$+1$&$+1$&$+1$&$+1$&$+1$&$+1$\\ % &$+1$&$+1$&$+1$&$+1$\\
$+1$&$+1$&$+1$&$+1$&$+1$&$+1$&{\color{green}$\bf{}+1$}&{\color{red}$\bf{}-1$}\\ % &$+1$&$+1$&$+1$&$-1$\\
$-1$&$-1$&$-1$&$-1$&$-1$&$-1$&$-1$&$-1$\\ % &$+1$&$+1$&$+1$&$+1$\\
$-1$&$-1$&$-1$&$-1$&$-1$&$-1$&{\color{green}$\bf{}-1$}&{\color{red}$\bf{}+1$}\\ % &$+1$&$+1$&$+1$&$-1$\\
$+1$&$+1$&$+1$&$+1$&$+1$&$+1$&$+1$&$+1$\\ % &$+1$&$+1$&$+1$&$+1$\\
$+1$&$+1$&$+1$&$+1$&$+1$&$+1$&{\color{green}$\bf{}+1$}&{\color{red}$\bf{}-1$}\\ % &$+1$&$+1$&$+1$&$-1$\\
$-1$&$-1$&$-1$&$-1$&$-1$&$-1$&$-1$&$-1$\\ % &$+1$&$+1$&$+1$&$+1$\\
$+1$&$+1$&$+1$&$+1$&$+1$&$+1$&$+1$&$+1$\\ % &$+1$&$+1$&$+1$&$+1$\\
$-1$&$-1$&$-1$&$-1$&$-1$&$-1$&{\color{green}$\bf{}-1$}&{\color{red}$\bf{}+1$}\\ % &$+1$&$+1$&$+1$&$-1$\\
$-1$&$-1$&$-1$&$-1$&$-1$&$-1$&$-1$&$-1$\\ % &$+1$&$+1$&$+1$&$+1$\\
$-1$&$-1$&$-1$&$-1$&$-1$&$-1$&{\color{green}$\bf{}-1$}&{\color{red}$\bf{}+1$}\\ % &$+1$&$+1$&$+1$&$-1$\\
$-1$&$-1$&$-1$&$-1$&$-1$&$-1$&$-1$&$-1$\\ % &$+1$&$+1$&$+1$&$+1$\\
$-1$&$-1$&$-1$&$-1$&$-1$&$-1$&$-1$&$-1$\\ % &$+1$&$+1$&$+1$&$+1$\\
$-1$&$-1$&$-1$&$-1$&$-1$&$-1$&$-1$&$-1$\\ % &$+1$&$+1$&$+1$&$+1$\\
$+1$&$+1$&$+1$&$+1$&$+1$&$+1$&{\color{green}$\bf{}+1$}&{\color{red}$\bf{}-1$}\\ % &$+1$&$+1$&$+1$&$-1$\\
$-1$&$-1$&$-1$&$-1$&$-1$&$-1$&{\color{green}$\bf{}-1$}&{\color{red}$\bf{}+1$}\\ % &$+1$&$+1$&$+1$&$-1$\\
$-1$&$-1$&$-1$&$-1$&$-1$&$-1$&$-1$&$-1$\\ % &$+1$&$+1$&$+1$&$+1$\\
$-1$&$-1$&$-1$&$-1$&$-1$&$-1$&{\color{green}$\bf{}-1$}&{\color{red}$\bf{}+1$}\\ % &$+1$&$+1$&$+1$&$-1$\\
$-1$&$-1$&$-1$&$-1$&$-1$&$-1$&$-1$&$-1$\\ % &$+1$&$+1$&$+1$&$+1$\\
$+1$&$+1$&$+1$&$+1$&$+1$&$+1$&{\color{green}$\bf{}+1$}&{\color{red}$\bf{}-1$}\\ % &$+1$&$+1$&$+1$&$-1$\\
$-1$&$-1$&$-1$&$-1$&$-1$&$-1$&{\color{green}$\bf{}-1$}&{\color{red}$\bf{}+1$}\\ % &$+1$&$+1$&$+1$&$-1$\\
\hline\hline
\end{tabular}
\end{center}
\caption{(Color online) 20 Counterfactual assignments of contextual (bold) and noncontextual values,
and the associated joint values, rendering an approximation $2.95$ for Tsirel'son's maximal quantum bound
$2\sqrt{2}$ for the CHSH sum.}
\label{2011-enough-t3}
\end{table}

With regards to Kochen-Specker type configurations~\cite{kochen1,cabello-96}
with no two-valued state,
any co-existing set of observables (associated with the configuration)
must breach noncontextuality at least once.
Other  Kochen-Specker type configurations~\cite{kochen1,svozil-ql,CalHerSvo}
still allowing two-valued states,
albeit an insufficient number for a homeomorphic embedding into Boolean algebras,
might still require contextual value assignments for quantum statistical reasons;
but this question remains unsolved at present.

In summary, several concrete, quantitative examples of contextual assignments
for co-existing complementary -- and thus strictly counterfactual -- observables
have been given.
The amount of noncontextuality can be characterized quantitatively by
the required relative amount of contextual assignments versus noncontextual ones
reproducing quantum mechanical predictions;
or, alternatively,  by
the required relative amount of contextual assignment versus all assignments.
One may thus consider the average number of contextual assignments per quantum as a criterion.

With regard to the above criteria, as could be expected, Kochen-Specker type configurations
require assignments which violate noncontextuality for every single quantum, whereas Boole-Bell-type
configurations, such as CHSH, would still allow occasional noncontextual assignments.
In this sense, Kochen-Specker-type arguments violate noncontextuality stronger than Boole-Bell-type ones.

These considerations are relevant under the assumption that contextuality is a viable concept
for explaining the
experiments~\cite{hasegawa:230401,cabello:210401,Bartosik-09,PhysRevLett.103.160405,kirch-09}.
As I have argued elsewhere~\cite{svozil-2004-analog,svozil-2006-omni,svozil:040102,svozil_2010-pc09},
this might not be the case; at least contextuality might not be a necessary quantum feature.
In particular the abandonment of quantum omniscience, in the sense that a quantum system can carry information
about its state with regard to only a {\em single} context~\cite{zeil-99},
in conjunction with a {\em context translation principle}~\cite{svozil-2003-garda,svozil-2008-ql}, thereby
effectively introducing stochasticity in the case
of a mismatch of preparation and measurement context,
might be an alternative approach to the quantum phenomena.

%\bibliography{svozil}


\end{document}